\documentclass[a4paper]{jpconf}
\usepackage{graphicx}

\usepackage{graphicx}
\usepackage{amssymb}
\usepackage{amsmath}    
\usepackage{txfonts}
\usepackage{xspace}

\newcommand{\Chandra}{\textit{Chandra}\xspace}
\newcommand{\Fermi}{\textit{Fermi}\xspace}

\newcommand{\He}{H.E.S.S.\xspace}
\newcommand{\Ve}{VERITAS\xspace}

\newcommand{\Ma}{MAGIC\xspace}

\newcommand{\M}{M\,87\xspace}

\begin{document}

\title{Probing the origin of VHE emission from \M with MWL observations in 2010}

\author{M. Raue$^{1}$,
L.~Stawarz$^{2}$,
D.~Mazin$^{3}$,
P.~Colin$^{4}$,
C.~M.~Hui$^{5}$,
M.~Beilicke$^{6}$,
R.C.~Walker$^{7}$\\
for the H.E.S.S., MAGIC, VERITAS, and Fermi/LAT Collaborations and the \M MWL Monitoring Teams}

\address{%
$^{1}$Institut f\"ur Experimentalphysik, Universit\"at Hamburg, Hamburg, Germany\\
$^{2}${Obserwatorium Astronomiczne, Uniwersytet Jagiello{\'n}ski, ul. Orla 171, 30-244 Krak{\'o}w, Poland}\\
$^{3}${IFAE, Edifici Cn., Campus UAB, E-08193 Bellaterra, Spain}\\
$^{4}${Max-Planck-Institut f\"ur Physik, D-80805 M\"unchen, Germany}\\
$^{5}${Department of Physics and Astronomy, University of Utah, Salt Lake City, UT 84112, USA}\\
$^{6}${Department of Physics, Washington University, St. Louis, MO 63130, USA}\\
$^{7}${National Radio Astronomy Observatory (NRAO), Socorro, NM 87801, USA}
}

\ead{martin.raue@desy.de}

\begin{abstract}
The large majority of extragalactic very high energy (VHE; $E>$100 GeV) sources belongs to the class of active galactic nuclei (AGN), in particular the BL Lac sub-class. AGNs are characterized by an extremely bright and compact emission region, powered by a super-massive black hole (SMBH) and an accretion disk, and relativistic outflows (jets) detected all across the electro-magnetic spectrum. In BL Lac sources the jet axis is oriented close to the line of sight, giving rise to a relativistic boosting of the emission. In radio galaxies, on the other hand, the jet makes a larger angle to the line of sight allowing to resolve the central core and the jet in great details. The giant radio galaxy \M with its proximity (16\,Mpc) and its very massive black hole ($(3-6) \times10^9 M_\odot$) provides a unique laboratory to investigate VHE emission in such objects and thereby probe particle acceleration to relativistic energies near SMBH and in jets. \M has been established as a VHE emitter since 2005. The VHE emission displays strong variability on time-scales as short as a day. It has been subject of a large joint VHE and multi-wavelength (MWL) monitoring campaign in 2008, where a rise in the 43\,GHz VLBA radio emission of the innermost region (core) was found to coincide with a flaring activity at VHE. This had been interpreted as a strong indication that the VHE emission is produced in the direct vicinity of the SMBH black hole. In 2010 again a flare at VHE was detected triggering further MWL observations with the VLBA, Chandra, and other instruments. At the same time \M was also observed with the Fermi/LAT telescope at GeV energies and the European VLBI Network (EVN). In this contribution preliminary results from the campaign will be presented.
\end{abstract}

\section{Introduction}

\begin{figure}[tb]
\centering
\includegraphics[width=0.5\textwidth]{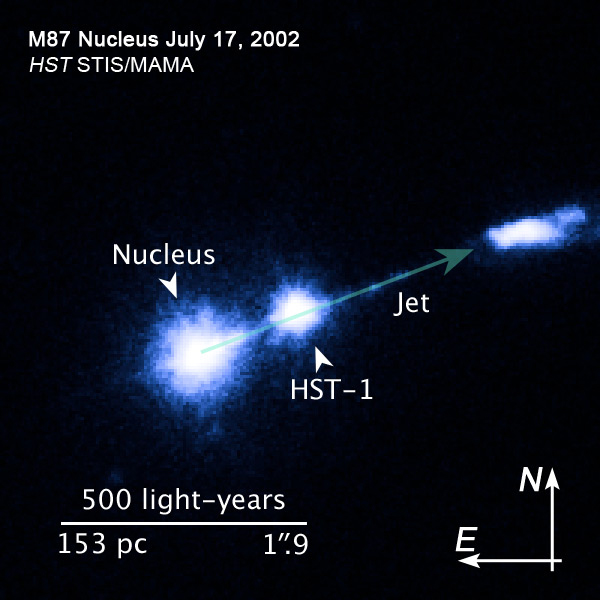}
\caption{Hubble space telescope (HST) image of M\,87 \cite{madrid:2009a}. \textit{(Illustration Credit: NASA, ESA, and Z. Levay (STScI); Credit: NASA, ESA, and J. Madrid (McMaster University))} \label{Fig:M87HST}}
\end{figure}

The giant radio galaxy \M provides a unique environment to study relativistic plasma outflows and the surrounding of super-massive black holes (SMBH). Its prominent jet \cite{curtis:1918a:m87jet} is resolved from radio to X-rays displaying complex structures (knots, diffuse emission; \cite{perlman:1999a,perlman:2001a}), strong variability \cite{harris:2003a,harris:2006a}, and super-luminal motion \cite{biretta:1999a,cheung:2007a} (Fig.~\ref{Fig:M87HST}). With its proximity ($16.7 \pm 0.2$\,Mpc; \cite{mei:2007a}) and its very massive black hole of $M_{\rm BH} \simeq (3-6) \times 10^{9}$\,M$_\odot$ \cite{macchetto:1997a,gebhardt:2009a}
high resolution radio observations enables one to directly probe structures with sizes down to $<200$ Schwarzschild radii.

\M has been established as a very high energy (VHE; $E>100$\,GeV) emitter since 2005 \cite{aharonian:2003b,aharonian:2006:hess:m87:science}. It is one of only 4 extragalactic VHE sources belonging to the class of radio galaxies, for which only weak or moderate beaming of the emission is expected. \M shows strong variability at VHE with time-scales of the order of days \cite{aharonian:2006:hess:m87:science,albert:2008:magic:m87,acciari:2009b:m87joinedcampaign:science}. This indicates a compact emission region $< 5 \times 10^{15} \delta$\,cm ($\delta$: bulk Doppler factor of the emitting plasma), corresponding to only a few tens of Schwarzschild radii. At GeV energies \M has recently been detected by \Fermi/LAT \cite{abdo:2009:fermi:lat:m87}.

The angular resolution of ground based VHE instruments\footnote{Typical $\sim$0.1\,deg per event, corresponding to $\sim30$\,kpc projected size.} does not allow for a direct determination of the origin of the VHE emission in the inner kpc structures. To further investigate the location of the VHE emission site and the associated production mechanisms variability studies and the search for correlations with other wavelengths need to be utilized (e.g. \cite{acciari:2009b:m87joinedcampaign:science}).

Up to now, three episodes of enhanced VHE activity have been detected from \M. The first one, detected in 2005 \cite{aharonian:2006:hess:m87:science}, occurred during an extreme radio/optical/X-ray outburst of the jet feature HST-1 \cite{harris:2003a,harris:2006a}, which has been discussed as possible emission site for the VHE emission (e.g. \cite{stawarz:2006a,cheung:2007a,harris:2009a}). During the second flaring episode, detected in 2008, HST-1 was in a low flux state, but radio measurements showed a flux increase of the core region within a few hundred Schwarzschild radii of the SMBH, suggesting the direct vicinity of the SMBH as the origin of the VHE emission \cite{acciari:2009b:m87joinedcampaign:science}. This conclusion was further supported by the detection of an enhanced X-ray flux from the core region by \Chandra.

The third episode of increased VHE activity occurred in 2010 during a joint VHE monitoring campaign by \Ma and \Ve. The detection of the high state \cite{mariotti:2010:magic:m87:atel,ong:2010a:m87:veritas:magic:flare:atel} triggered further MWL observations by the VLBA, \Chandra, and other instruments. Preliminary results from the campaign are presented in this paper, while the final campaign results will be reported in an upcoming publication.

\section{The 2010 VHE flare}

\begin{figure}[tb]
\centering
\includegraphics[width=0.9\textwidth]{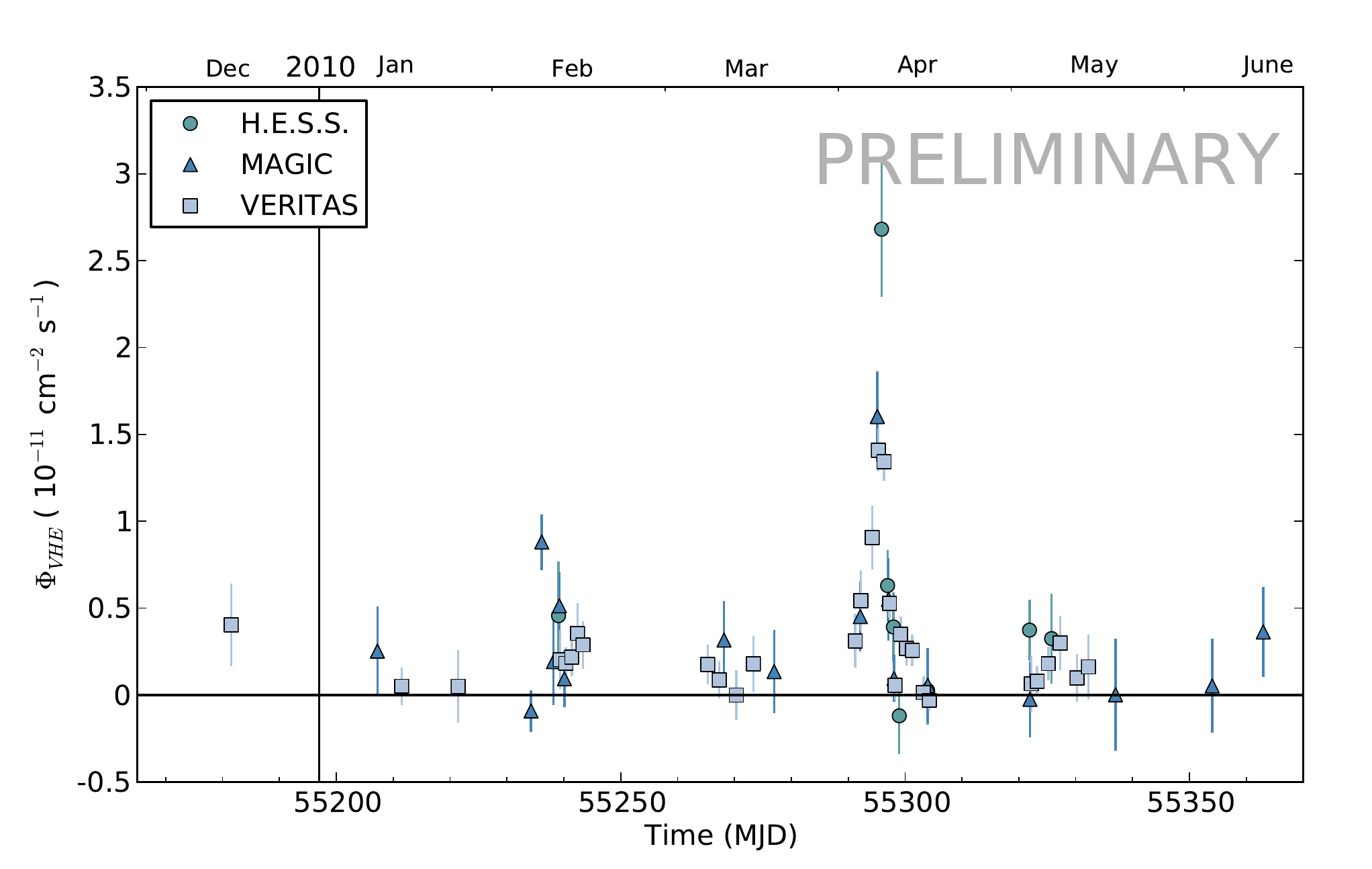}
\caption{Preliminary VHE lightcurve from the 2010 monitoring campaign on \M as recorded by \He, \Ma, and \Ve. The integral flux above an energy of 300\,GeV is shown. \label{Fig:VHELightcurve}}
\end{figure}

The preliminary combined VHE lightcurve of the 2010 monitoring campaign is shown in Fig.~\ref{Fig:VHELightcurve}. During the campaign, two VHE flares have been detected \cite{mariotti:2010:magic:m87:atel,ong:2010a:m87:veritas:magic:flare:atel}: The first episode took place in Feb. 2010 where a single night of increased activity was detected by \Ma. Follow up observations did not reveal further activity. The second episode took place in Apr. 2010 and showed a pronounced VHE flare detected by several instruments triggering further MWL observations.
The VHE activity of this second flaring episode is concentrated in a single observation period between MJD 55290 and MJD 55305 ($\sim$15\,days). This time-period is exceptionally well covered by observations with 21 pointings by different VHE instruments resulting in an observation almost every night. The detected flare displays a smooth rise and decay in flux with a peak around MJD 55296 (April 9-10, 2010; see Fig.~\ref{Fig:VHELightcurve}). In general, during nights with (quasi) simultaneous observations by different instruments, the measured fluxes are found to be in excellent agreement.

In comparison to previous VHE flares detected in 2005 and 2008, the 2010 flare shows similar time-scales and peak flux levels, but the overall variability pattern is somewhat different to what has been observed before (more extended periods of flaring activities with several flux maxima), though the statistics and the sampling of the previous VHE flares limit a definitive conclusion. From the VHE long-term lightcurve spanning from 2004 to 2010 the duty cycle for VHE flares is estimated to be $\sim14-4$\% (depending on the assumed threshold flux defining a VHE high state).

\section{2010 MWL observations}

The discovery of a VHE high state in April 2010 triggered further MWL observations.
\Chandra started observing $\sim2$\,days after the peak VHE flux had been detected performing five pointings  spaced in intervals between 1.5 and 3 days (5ks each). A second set of four observations was taken starting about two weeks later. HST-1 was found in a low flux state while the core showed an increase in X-ray flux in the first observation that followed the VHE flare.

Five VLBA 43\,GHz observations where taken in 2010, three monitoring observations and two additional observations following the detection of the VHE high state. No enhanced radio flux from the core region was detected, contrary to what has been observed during the 2008 VHE outburst.

During the 2010 campaign, further MWL data were taken by the Liverpool Telescope, the EVN, and the VLBA, and \M was continually monitored at MeV/GeV energies with the Fermi-LAT. Results from these observations will be presented in the upcoming publication.

 \ack
{\small
 The author acknowledges support by the LEXI program of the state of Hamburg, Germany.
--- The \He Collaboration acknowledges support of the Namibian authorities and of the University of Namibia
in facilitating the construction and operation of H.E.S.S., as is the support by the German Ministry for Education and
Research (BMBF), the Max Planck Society, the French Ministry for Research,
the CNRS-IN2P3 and the Astroparticle Interdisciplinary Programme of the
CNRS, the U.K. Science and Technology Facilities Council (STFC),
the IPNP of the Charles University, the Polish Ministry of Science and 
Higher Education, the South African Department of
Science and Technology and National Research Foundation, and by the
University of Namibia. We appreciate the excellent work of the technical
support staff in Berlin, Durham, Hamburg, Heidelberg, Palaiseau, Paris,
Saclay, and in Namibia in the construction and operation of the
equipment.
--- The MAGIC Collaboration thanks the Instituto de Astrof'sica de Canarias for the excellent working conditions at the Observatorio del Roque de los Muchachos in La Palma. The support of the German BMBF and MPG, the Italian INFN and Spanish MICINN is gratefully acknowledged.
--- The VERITAS Collaboration acknowledges support from the U.S. Department of Energy, the U.S. National Science Foundation and the Smithsonian Institution, by NSERC in Canada, by Science Foundation Ireland, and by STFC in the UK.
--- The $Fermi$ LAT Collaboration acknowledges support from a number of agencies and institutes for both development and the operation of the LAT as well as scientific data analysis. These include NASA and DOE in the United States, CEA/Irfu and IN2P3/CNRS in France, ASI and INFN in Italy, MEXT, KEK, and JAXA in Japan, and the K.~A.~Wallenberg Foundation, the Swedish Research Council and the National Space Board in Sweden. Additional support from INAF in Italy and CNES in France for science analysis during the operations phase is also gratefully acknowledged.
--- The Very Long Baseline Array is operated by the
National Radio Astronomy Observatory, a facility of the NSF, operated
under cooperative agreement by Associated Universities, Inc.
--- The European VLBI Network is a joint facility of European, Chinese, South African and other radio astronomy institutes funded by their national research councils. This effort is supported by the European Community Framework Programme 7, Advanced Radio Astronomy in Europe, grant agreement No. 227290.
--- This research has made use of data from the MOJAVE database that is maintained by the MOJAVE team \cite{lister:2009a}.
--- This research has made use of NASA's Astrophysics Data System.
}


\section*{References}
\def\Journal#1#2#3#4{{#4}, {#1}, {#2}, #3}
\def\NAT{Nature}
\def\AAA{A\&A}
\def\ApJ{ApJ}
\def\AJ{Astronom. Journal}
\def\Aph{Astropart. Phys.}
\def\ApJS{ApJSS}
\def\MNRAS{MNRAS}
\def\NIM{Nucl. Instrum. Methods}
\def\NIMA{Nucl. Instrum. Methods A}
\def\aj{AJ}%
\def\actaa{Acta Astron.}%
\def\araa{ARA\&A}%
\def\apj{ApJ}%
\def\apjl{ApJ}%
\def\apjs{ApJS}%
\def\ao{Appl.~Opt.}%
\def\apss{Ap\&SS}%
\def\aap{A\&A}%
\def\aapr{A\&A~Rev.}%
\def\aaps{A\&AS}%
\def\azh{AZh}%
\def\baas{BAAS}%
\def\bac{Bull. astr. Inst. Czechosl.}%
\def\caa{Chinese Astron. Astrophys.}%
\def\cjaa{Chinese J. Astron. Astrophys.}%
\def\icarus{Icarus}%
\def\jcap{J. Cosmology Astropart. Phys.}%
\def\jrasc{JRASC}%
\def\mnras{MNRAS}%
\def\memras{MmRAS}%
\def\na{New A}%
\def\nar{New A Rev.}%
\def\pasa{PASA}%
\def\pra{Phys.~Rev.~A}%
\def\prb{Phys.~Rev.~B}%
\def\prc{Phys.~Rev.~C}%
\def\prd{Phys.~Rev.~D}%
\def\pre{Phys.~Rev.~E}%
\def\prl{Phys.~Rev.~Lett.}%
\def\pasp{PASP}%
\def\pasj{PASJ}%
\def\qjras{QJRAS}%
\def\rmxaa{Rev. Mexicana Astron. Astrofis.}%
\def\skytel{S\&T}%
\def\solphys{Sol.~Phys.}%
\def\sovast{Soviet~Ast.}%
\def\ssr{Space~Sci.~Rev.}%
\def\zap{ZAp}%
\def\nat{Nature}%
\def\iaucirc{IAU~Circ.}%
\def\aplett{Astrophys.~Lett.}%
\def\apspr{Astrophys.~Space~Phys.~Res.}%
\def\bain{Bull.~Astron.~Inst.~Netherlands}%
\def\fcp{Fund.~Cosmic~Phys.}%
\def\gca{Geochim.~Cosmochim.~Acta}%
\def\grl{Geophys.~Res.~Lett.}%
\def\jcp{J.~Chem.~Phys.}%
\def\jgr{J.~Geophys.~Res.}%
\def\jqsrt{J.~Quant.~Spec.~Radiat.~Transf.}%
\def\memsai{Mem.~Soc.~Astron.~Italiana}%
\def\nphysa{Nucl.~Phys.~A}%
\def\physrep{Phys.~Rep.}%
\def\physscr{Phys.~Scr}%
\def\planss{Planet.~Space~Sci.}%
\def\procspie{Proc.~SPIE}%
          
\providecommand{\newblock}{}

\end{document}